\documentclass[sigconf,nonacm]{acmart}

\usepackage{booktabs} 
\usepackage{amsmath,amssymb,amsfonts}
\usepackage{algorithm}
\usepackage{algorithmic}
\usepackage{float}
\usepackage{setspace}
\usepackage{graphicx}
\usepackage{textcomp}
\usepackage{xcolor}
\usepackage{subfigure}
\usepackage{verbatim}
\usepackage{enumitem}

\newtheorem{Def}{Definition}[section]

\AtBeginDocument{%
  \providecommand\BibTeX{{%
    \normalfont B\kern-0.5em{\scshape i\kern-0.25em b}\kern-0.8em\TeX}}}

\setcopyright{acmcopyright}
\copyrightyear{2018}
\acmYear{2018}
\acmDOI{10.1145/1122445.1122456}

\begin{document}

\title{Trading Location Data with Bounded Personalized Privacy Loss}

\author{Shuyuan Zheng, Yang Cao, Masatoshi Yoshikawa}
\email{caryzheng@db.soc.i.kyoto-u.ac.jp,
	   yang@i.kyoto-u.ac.jp,
	   yoshikawa@i.kyoto-u.ac.jp}
\affiliation{%
  \institution{Kyoto University, Kyoto, Japan}
}


\begin{abstract}
As personal data have been the new oil of the digital era, there is a growing trend perceiving personal data as a commodity. Although some people are willing to trade their personal data for money, they might still expect limited privacy loss, and the maximum tolerable privacy loss varies with each individual. In this paper, we propose a framework that enables individuals to trade their personal data with bounded personalized privacy loss, which raises technical challenges in the aspects of \textit{budget allocation} and \textit{arbitrage-freeness}. To deal with those challenges, we propose two arbitrage-free trading mechanisms with different advantages.
\end{abstract}




\maketitle

\section{Introduction}

Personal data, the new oil of the digital era, are extraordinarily valuable for individuals and organizations to discover knowledge and improve products or services. However, data owners' personal data have been exploited without appropriate compensations. Very recently, there is a growing trend towards \textit{personal data trading} perceiving personal data as a commodity, which meets the demand of both data buyers and data owners. Some companies also consider personal data trading platform that connects data owners and buyers directly as a new business model.

Several studies \cite{ghosh2015selling}\cite{koutris2013toward}\cite{koutris2015query}\cite{li2014theory}\cite{niu2018unlocking} in the
literature investigated privacy preserving query-based data trading.
There are three parties in the data trading: data owners, data buyers, and a market maker. 
Data owners contribute their personal data and get monetary compensations from the market maker in return. 
Data buyers request queries over the data and purchase perturbed query answers, i.e., noisy versions of aggregate statistical results where some random noises are injected. 
The market maker acts as a trustworthy intermediary between data owners and data buyers, in charge of computing perturbed query answers, pricing data for data buyers and compensating data owners. 
A major challenge in the line of works is how to determine the price of data. 
A seminal work of Li et al. \cite{li2014theory} made the connection between privacy loss and the price of data by pricing function.
Specifically, data owners who suffer higher privacy loss (when less noises are injected into the query answer) should be compensated with more money.
They proposed an important property of pricing function: \textit{arbitrage-freeness}, which means the consistency of a set of priced queries. 
Intuitively, a buyer should not obtain the answer to a query more cheaply by deriving the answer from a less expensive set of queries. 

However, there are several insufficiencies in such a marketplace where each data owner contributes their location data.
First, data owners should be able to bound their privacy loss given the high sensitivity of locations. 
In the traditional data marketplace \cite{li2014theory}, a buyer can purchase raw data and data owners cannot control the upper bound of their privacy loss.
Second, the existing studies \cite{ghosh2015selling}\cite{li2014theory}\cite{niu2018unlocking} guarantee uniform privacy loss from each data owner; however, it is more natural that different data owners have quite diverse expectations on tolerable privacy loss \cite{jorgensen2015conservative}. 

We proposed a new trading framework where each owner is enabled to set her own personalized bound of privacy loss. In order to achieve better utility for data buyers under the constraint of arbitrage-freeness, we proposed a budget allocation algorithm called $N$-Grouping for the Sample mechanism \cite{jorgensen2015conservative} which achieves personalized differential privacy \cite{jorgensen2015conservative}. To easily design arbitrage-free pricing functions, we proposed a theorem presenting the sufficient conditions in which the patterned utility function of some perturbation mechanism will result in an arbitrage-free pricing function. 

\section{Problem Formulation}
In this section, we introduce the basic settings in our framework.

\subsubsection*{Privacy Definition}
As we are going to trade personal data, first we should define the privacy metric in our framework. To preserve data owners' privacy, the true query answer cannot be returned to the buyer and the market maker should perturb the answer by some perturbation mechanism $\mathcal{M}$ achieving a formal privacy standard. We follow the setting of \cite{li2014theory} to use differential privacy as privacy metric.

In $\epsilon$-differential privacy, the value of privacy protection metric $\epsilon$ is uniform for all data owners. However, the setting that data owners are allowed to set their maximum tolerable privacy losses requires personalized privacy protection level. In personalized differential privacy (PDP) \cite{jorgensen2015conservative}, a privacy loss $\epsilon_i$ is a personalized privacy protection metric, and a privacy specification $\Phi$ is a mapping from data owners $u_i$ to their privacy losses $\epsilon_i$. 

\begin{Def}[Personalized Differential Privacy \cite{jorgensen2015conservative}]
\label{PDP}
Given a privacy specification $\Phi=\{(u_i,\epsilon_i)|u_i \in User\}$, a perturbation mechanism $\mathcal{M}: \mathcal{D} \rightarrow \mathcal{R}^d$ ($d \in Z^+$) satisfies $\Phi$-personalized differential privacy, if for any pair of neighboring databases $D, D'\subset D$, with $D \stackrel{u_i}{\sim} D^\prime$, and for any possible output $o\in \mathcal{R}^d$, we have:
\begin{equation*}
Pr[\mathcal{M}(D)=o] \leq e^{\epsilon_i} Pr[\mathcal{M}(D')=o]
\end{equation*}
Two databases $D, D'\subset \mathcal{D}$ are neighboring if $D'$ can be derived from $D$ by replacing one data point (row) with another, denoted as $D\sim D'$. We write $D \stackrel{u_i}{\sim} D^\prime$ to denote $D$ and $D'$ are neighboring and differ on the value of data point $p_i$ corresponding to the data owner $u_i$.
\end{Def}

\subsubsection*{Participants in Data Marketplace}
There are three parties of participants in our marketplace: data owners, data buyers, and a market maker. Each party has its own interests and goals.

\textbf{Data owners} are the source of location data stored in the marketplace. A database $D$ consists of $n$ data points, where each data point $p_i$ corresponds to a unique data owner $u_i$ and its value is a location $l$. Each data owner $u_i$ should specify her maximum tolerable privacy loss $\hat{\epsilon}_i$ leaked totally for query answers over a database $D$. 

\textbf{Data buyers} can request a histogram query $Q^k$ (where $k \in Z^+$ is an order number) over the database $D$ with a specific utility $v^k$, and get a perturbed query answer $q^k=Q^k(D,v^k)$. The perturbed query answer $q^k$ is a vector where each element $q^k_i$ represents the sum of data owners in the location $l_i$. 

\textbf{The market maker} acts as a trustworthy profit-making intermediary between data owners and data buyers: 
for the sake of data owners' privacy, the maker is in charge of perturbing the histogram query by some perturbation mechanism $\mathcal{M}$ which satisfies PDP (i.e., Definition \ref{PDP}); 
for buyers, the maker should guarantee that the variance of each element in the perturbed query answer $q^k=Q^k(D,v^k)$ is no more than $v^k$, i.e., $\forall i, Var(q^k_i)\leq v^k$.

As a profit-making intermediary, the maker sells histogram queries to buyers and compensates data owners according to their privacy losses. Each data owner $u_i$ should make a contract with the market maker that the latter compensates the former according to a compensation function $\mu_i$. Let $\epsilon_i^k$ be the data owner $u_i$'s privacy loss due to the query answer $q^k$. Each owner's compensation $\mu_i(\epsilon_i^k)=c_i \cdot \epsilon_i^k$ depends on her privacy loss $\epsilon_i^k$ where $c_i>0$ is a constant compensation rate stipulated in the contract. We also define our utility function as $v=U([\epsilon_1,...,\epsilon_n])$ which takes as input a vector of privacy losses (or budgets) and outputs the histogram variance $v$ mentioned above. According to the Composition theorem of PDP \cite{jorgensen2015conservative}, in order to achieve PDP, for each data owner $u_i$, we should make sure: $\sum_k \epsilon^k_i \leq \hat{\epsilon}_i$, which means for each data owner $u_i$, the sum of privacy losses should be no more than her maximum tolerable privacy loss. For the sake of data owners, we try to make full use of their maximum tolerable privacy losses so that they can gain more compensations.

The maker always charges a brokerage fee $r\cdot \sum_i \mu_i(\epsilon^k_i)$ as her profit, where $r\geq 0$ is a constant rate for all queries. However, we set the query price $\pi^k=\Pi(v^k)$ determined only by the histogram variance $v^k$. The maker should guarantee that the pricing function $\Pi$ is arbitrage-free to prevent arbitrage behaviors. Arbitrage-freeness intuitively requires that, the buyer should not obtain the answer to a query more cheaply by deriving this answer from a less expensive set of query answers. On the other hand, because the price of a query answer should consist of all the compensations to data owners and the maker's profit, we should make sure that the price $\pi^k=\Pi(v^k)=(1+r)\cdot \sum_i \mu_i(\epsilon^k_i)$ for all $k$. For the sake of data buyers, we endeavor to provide more choices of the value of $v^k$ and lower the $\Pi(v^k)$ by improving the perturbation mechanism. 

\begin{Def}[Arbitrage-freeness \cite{li2014theory}]
A pricing function $\pi=\Pi(v)$ ($v>0$) is arbitrage-free if: for every multiset $V=\{v_1,...,v_m\}$ where $m \in Z^+$, if there exists $a_1,...,a_m$ such that $\sum_{j=1}^m a_j=1$ and $\sum_{j=1}^m a_j^2v_j\leq v$, then 
$\Pi(v)\leq \sum_{j=1}^m\Pi(v_j)$. 
\end{Def}

\section{Data Trading Framework}
In this section, we give an overview of our trading framework. We summarize some important notations in Table \ref{table:notations}.
\begin{table}[t]
\caption{Summary of Notations}
\begin{spacing}{1.25}
	\begin{tabular}{|c||c|}
	\hline
		Notation & Description\\	
	\hline
		$u_i$ & data owner $i$\\
	\hline
		$n$ & total number of users, or database size\\
	\hline
		$k$ & order number of a query\\
	\hline
		$q^k$ & perturbed query answer of $Q^k$\\
	\hline
		$v^k$ & histogram variance of $q^k$\\
	\hline
		$\check{v}^k$ & minimum affordable variance for $Q^k$ \\
	\hline
		$\hat{\epsilon}_i$ & $u_i$'s maximum tolerable privacy loss\\
	\hline
		$\hat{\epsilon}_i^k$ & $u_i$'s remaining tolerable privacy loss for $Q^k$\\
	\hline
		$\bar{\epsilon}_i^k$ & privacy budget for $Q^k$\\	
	\hline
		$\epsilon_i^k$ & $u_i$'s privacy loss for $q^k$\\
	\hline
		$\mu_i(\cdot)$ & $u_i$'s compensation function\\
	\hline
		$\mathcal{M}$ & perturbation mechanism\\
	\hline	
		$v=U_{\mathcal{M}}([\epsilon_1,...,\epsilon_n])$ & utility function of $\mathcal{M}$\\ 
	\hline
		$v=U^{\mathbf{\rho}}_{\mathcal{M}}(\epsilon_{base})$ & patterned utility function of $\mathcal{M}$ in $\mathbf{\rho}$-pattern\\
	\hline
		$\pi=\Pi^{\mathbf{\rho}}_{\mathcal{M}}(v)$ & pricing function of $\mathcal{M}$ in $\mathbf{\rho}$-pattern\\
	\hline
\end{tabular}
\end{spacing}
\label{table:notations}
\end{table}

\begin{figure}[ht]
\centerline{\includegraphics[scale=0.25]{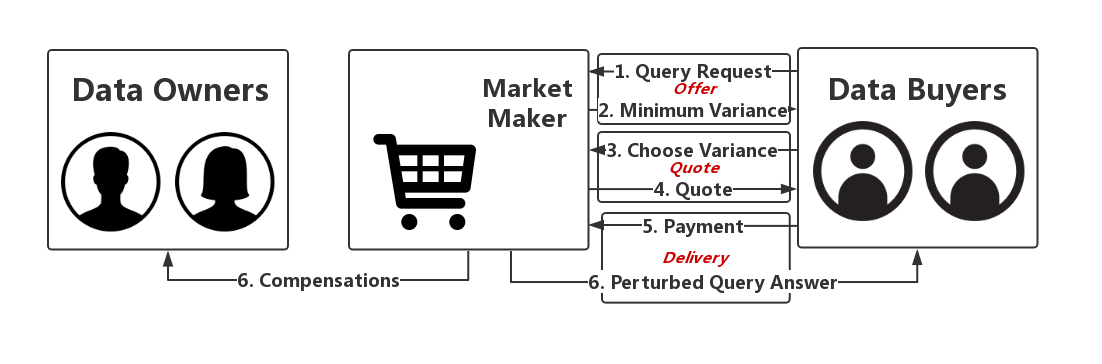}}
\caption{Trading Framework}
\label{trading_framework}
\end{figure}

Figure \ref{trading_framework} illustrates our trading framework containing three phases: \textit{Offer}, \textit{Quote}, and \textit{Delivery}. As depicted in Alg. \ref{alg:framework}, for each query $Q^k$, the transaction follows those phases; the maker sells query answers $q^k$ until some data owner's privacy loss reaches her maximum tolerable privacy loss, i.e., $\exists u_i, \sum_{j=1}^k \epsilon^j_i = \hat{\epsilon}_i$.

\textbf{Offer: } 
In the phase of Offer, at the beginning, the buyer requests a query $Q^k$ over the database $D$. Then, the market maker executes the process of budget allocation (\textit{BudgetAlloc}). That is, given remaining tolerable privacy losses $[\hat{\epsilon}^k_1,...,\hat{\epsilon}^k_n]$, the maker calculates privacy budget $\bar{\epsilon}^k_i$ for each $u_i$ by some budget allocation algorithm. Finally, the maker inputs the budgets $[\bar{\epsilon}^k_1,...,\bar{\epsilon}^k_n]$ into the utility function $U_{\mathcal{M}}$ to return the minimum affordable variance $\check{v}^k$ for $Q^k$.

\textbf{Quote: }
Given the $\check{v}^k$, the buyer should choose a histogram variance $v^k\geq\check{v}^k$. Then, the maker quotes a price $\pi^k$ for the query answer $q^k=Q^k(D,v^k)$. Because we should guarantee that the price is also equal to the sum of all the compensations and the maker's profit, i.e., $\pi^k = (1+r)\cdot \sum_i \mu_i(\epsilon^k_i)$, the ideal way to design a pricing function is to establish a mapping from the set of $v$ to the set of $[\epsilon_1,...,\epsilon_n]$, which requires the existence of the inverse function ${U_{\mathcal{M}}}^{-1}$. To make ${U_{\mathcal{M}}}^{-1}$ existent, we introduce the notion \textit{Pattern} to "shape" privacy losses (or budgets) and constrain the utility function's domain. 

\begin{Def}[$\mathbf{\rho}$-Pattern]
A pattern is a vector $\mathbf{\mathbf{\rho}}=[\rho_1,...,\rho_n]$ where $0 \leq \rho_i \leq 1$ for all $i$ and $\max\limits_i\rho_i=1$. A vector $[\epsilon_1,...,\epsilon_n]$ is in $\mathbf{\rho}$-pattern, if $\frac{\epsilon_i}{\epsilon_{base}}=\rho_i$ for all $i$, where $\epsilon_i \geq 0$ for all $i$ and $\epsilon_{base}=\max\limits_j\epsilon_j>0$. $U_{\mathcal{M}}^{\mathbf{\rho}}$ is the patterned utility function in $\mathbf{\rho}$-pattern such that $U_{\mathcal{M}}^{\mathbf{\rho}}(\epsilon_{base})=U_{\mathcal{M}}([\rho_1\cdot\epsilon_{base},...,\rho_n\cdot\epsilon_{base}])$ for all $\epsilon_{base}>0$. $\Pi^{\mathbf{\rho}}_{\mathcal{M}}$ is the pricing function in $\mathbf{\rho}$-pattern such that $\Pi^{\mathbf{\rho}}_{\mathcal{M}}(v)=(1+r)\cdot \sum_i c_i \cdot \rho_i \cdot {U_{\mathcal{M}}^{\mathbf{\rho}}}^{-1}(v)$ for all $v>0$.
\end{Def}

If the domain of $v={U_{\mathcal{M}}}([\epsilon_1,...,\epsilon_n])$ is the set of vectors in $\mathbf{\rho}$-pattern, we can use a scalar $\epsilon_{base}=\max_i \epsilon_i$ to represent the vector $[\epsilon_1,...,\epsilon_n]$. Therefore, we establish a mapping from the set of $\epsilon_{base}$ to the set of $v$, i.e., $v=U_{\mathcal{M}}^{\mathbf{\rho}}(\epsilon_{base})$ (the patterned utility function in $\mathbf{\rho}$-pattern). Our budget allocation algorithm guarantees $U_{\mathcal{M}}^{\mathbf{\rho}}$ is a decreasing function and thus ${U_{\mathcal{M}}^{\mathbf{\rho}}}^{-1}$ exists. Then, the market maker can derive the price $\pi^k$ as follows:
\begin{equation*}
\pi^k=\Pi^{\mathbf{\rho}}_{\mathcal{M}}(v^k)=(1+r)\cdot \sum_i c_i \cdot \rho_i \cdot {U_{\mathcal{M}}^{\mathbf{\rho}}}^{-1}(v^k) =(1+r)\cdot \sum_i \mu_i(\epsilon^k_i)
\end{equation*}

As for arbitrage-freeness, intuitively, if the price decreases rapidly as the variance increases, it is more worthwhile to buy a query answer with high variance. Thus, to avoid arbitrage behaviors, we can find a pattern $\rho$ by budget allocation algorithm, such that $\epsilon_{base}$ decreases smoothly to some extent as $v^k$ increases and $\pi^k=\Pi^{\mathbf{\rho}}_{\mathcal{M}}(v^k)$ decreases slow enough, as implied mathematically by the second condition in Theorem \ref{Arbitrage-free}.

\textbf{Delivery: }
In the phase of Delivery, the buyer pays the price $\pi^k$ to the market maker and receives a perturbed query answer with the histogram variance $v^k$ in return, i.e., $q^k=Q^k(D,v^k)$. The maker calculates each owner's privacy loss $\epsilon^k_i$ for this variance by the inverse function ${U^{\mathbf{\rho}}_{\mathcal{M}}}^{-1}$, and then compensates data owners. The answer is perturbed by $\mathcal{M}$ (an instance of \textit{PerturbMech}) which achieves $\Phi^k$-PDP, where $\Phi^k=\{(u_i,\epsilon^k_i)|u_i \in User\}$.

\begin{algorithm}[h]
\algsetup{linenosize=\small} 
\small
\caption{Trading Data}
\label{alg:framework}
\begin{algorithmic}[1]
\REQUIRE required variance $v^k$
\ENSURE perturbed query answer $q^k$

\STATE $k\gets 1,[\hat{\epsilon}^k_1,...,\hat{\epsilon}^k_1]\gets [\hat{\epsilon}_1,...,\hat{\epsilon}_n]$;

\WHILE {\textbf{not} $\exists \hat{\epsilon}^k_i=0$}

\STATE $[\bar{\epsilon}^k_1,...,\bar{\epsilon}^k_n]\gets \textit{BudgetAlloc}([\hat{\epsilon}^k_1,...,\hat{\epsilon}^k_n])$;
\STATE $\check{v}^k \gets U_{\mathcal{M}}([\bar{\epsilon}^k_1,...,\bar{\epsilon}^k_n])$;

\IF{$v^k<\check{v}^k$}
	\STATE Reject the query;
\ELSE
	\STATE $\mathbf{\rho} \gets [\bar{\epsilon}^k_1,...,\bar{\epsilon}^k_n] / \max_i \bar{\epsilon}^k_i$;
	\STATE Quote the buyer $\Pi^{\mathbf{\rho}}_{\mathcal{M}}(v^k)$;
	\STATE Collect the buyer's payment;
	\STATE $[\epsilon^k_1,...,\epsilon^k_n] \gets {U_{\mathcal{M}}^{\mathbf{\rho}}}^{-1}(v^k)\cdot \mathbf{\rho}$;
	\STATE Compensate $\mu_i(\epsilon^k_i)$ to each data owner $u_i$;
	\STATE $q^k \gets \textit{PerturbMech}(D,[\epsilon^k_1,...,\epsilon^k_n])$;
	\STATE Return back to the buyer the perturbed query answer $q^k$;
	\STATE $[\hat{\epsilon}^{k+1}_1,...,\hat{\epsilon}^{k+1}_1]\gets [\hat{\epsilon}^k_1,...,\hat{\epsilon}^k_1] - [\epsilon^k_1,...,\epsilon^k_n]$;
	\STATE $k \gets k+1$;
\ENDIF
\ENDWHILE
\end{algorithmic}
\end{algorithm}

\section{Arbitrage-free Trading Mechanism}
In this section, we propose two arbitrage-free trading mechanisms by combining instances of two key modules in Alg. \ref{alg:framework}: \textit{BudgetAlloc} (budget allocation), and \textit{PerturbMech} (perturbation mechanism).

\subsubsection*{Laplace-base Mechanism: Laplace + Uniform}
In the Laplace-base mechanism, we use the Laplace mechanism for \textit{PerturbMech} and the Uniform algorithm for \textit{BudgetAlloc}. According to Theorem \ref{Laplace}, because the Laplace mechanism always controls the privacy protection level by the minimum of $[\bar{\epsilon}_1,...,\bar{\epsilon}_n]$, it is waste to allocate privacy budgets $\bar{\epsilon}_j > \min_i\{\bar{\epsilon}_i\}$. Thus, we use Uniform to allocate uniform privacy budgets for the Laplace mechanism. 

\begin{theorem}[Laplace Mechanism \cite{jorgensen2015conservative}]
\label{Laplace}
Given a function $f:\mathcal{D}\rightarrow\mathcal{R}^d$, a database $D$, and privacy budgets $[\bar{\epsilon}_1,...,\bar{\epsilon}_n]$, the Laplace mechanism which returns $f(D)+ \mathcal{Z}^d$ satisfies $\Phi$-PDP, where $\mathcal{Z}^d$ are random variables drawn from the Laplace distribution $Lap(\frac{\Delta_f}{\min_i\{\bar{\epsilon}_i\}})$ ($\Delta_{f}=\max\limits_{D \sim D^\prime}\left\|f(D)-f\left(D^{\prime}\right)\right\|_{1}$) and $\Phi=\{(u_i,\min_i\{\bar{\epsilon}_i\})|u_i \in User\}$.
\end{theorem}

\begin{Def}[Uniform]
Given a query $Q^k$, the Uniform algorithm allocates $\frac{1}{2}\min_i\{\hat{\epsilon}_i-\sum_{j=1}^{k-1} \epsilon^j_i\}$ to $\bar{\epsilon}^k_i$ for each $u_i$. 
\end{Def}

We can easily observe that $[\bar{\epsilon}^k_1,...,\bar{\epsilon}^k_n]$ allocated by Uniform is in $\mathbf{\rho}$-pattern where $\rho_i=1$ for all $i$, and the patterned utility function $U^{\mathbf{\rho}}_{Lap}(\epsilon_{base})=2\cdot (\frac{\Delta f}{\epsilon_{base}})^2$. Also, $\pi=\Pi^{\mathbf{\rho}}_{Lap}(v)$ is arbitrage-free. However, the Laplace-base mechanism cannot essentially achieve our goal, i.e., to enable each owner's privacy loss to be personalized.

\subsubsection*{Sample-base Mechanism: Sample + Grouping}
In order to personalize privacy loss for each data owner, we propose the Sample-base mechanism which is the combination of the Sample mechanism for \textit{PerturbMech} and the $N$-Grouping algorithm for \textit{BudgetAlloc}. According to Theorem \ref{Sample}, under the Sample mechanism, each $u_i$'s privacy loss $\epsilon^k_i$ can be personalized rather than only uniform. 

\begin{theorem}[Sample Mechanism \cite{jorgensen2015conservative}]
\label{Sample}
Given a function $f:\mathcal{D}\rightarrow\mathcal{R}^d$, a database $D$, and privacy budgets $[\bar{\epsilon}_1,...,\bar{\epsilon}_n]$, the Sample mechanism samples each data point $p_i \in D$ with probability $Pr_i=\frac{e^{\bar{\epsilon}_i}-1}{e^{\max_j\bar{\epsilon}_j}-1}$, and then executes the Laplace mechanism to return $f(\tilde{D})+Lap(\frac{\Delta_f}{\max_j \bar{\epsilon}_j})$ where $\Delta_{f}=\max _{\tilde{D} \sim \tilde{D}^\prime}\left\|f(\tilde{D})-f\left(\tilde{D}^{\prime}\right)\right\|_{1}$ and $\tilde{D}$ is the sampled database of $D$. The Sample mechanism satisfies $\Phi$-PDP privacy, where $\Phi=\{(u_i,\bar{\epsilon}_i)|u_i \in User\}$. 
\end{theorem}

The utility function $U_{Sam}([\epsilon_1,...,\epsilon_n])=\sum_i \frac{e^{\epsilon_i}-1}{\max_j e^{\epsilon_j}-1} \cdot (1-\frac{e^{\epsilon_i}-1}{\max_j e^{\epsilon_j}-1})+2\cdot (\frac{\Delta f}{\max_j\epsilon_j})^2$, and if its domain is constrained to be the set of vectors in $\mathbf{\rho}$-pattern, we can derive the patterned utility function as:
$
U^{\mathbf{\rho}}_{Sam}(\epsilon_{base})=\sum_j \frac{e^{\rho_j \epsilon_{base}}-1}{e^{\epsilon_{base}}-1} \cdot (1-\frac{e^{\rho_j \epsilon_{base}}-1}{e^{\epsilon_{base}}-1})+2\cdot (\frac{\Delta f}{\epsilon_{base}})^2$.
We can find some pattern $\mathbf{\rho}$ such that $U^{\mathbf{\rho}}_{Sam}$ is decreasing and therefore the inverse function ${U^{\mathbf{\rho}}_{Sam}}^{-1}$ exists. However, the pricing function $\Pi^{\mathbf{\rho}}_{Sam}(v) = (1+r)\cdot \sum_i c_i \cdot \rho_i \cdot {U^{\mathbf{\rho}}_{Sam}}^{-1}(v)\cdot =(1+r) \sum_i c_i \cdot \epsilon_i$ might be not arbitrage-free as depicted in Example \ref{arbitrage}. That is because, in some $\mathbf{\rho}$-pattern, $v=U^{\mathbf{\rho}}_{Sam}(\epsilon_{base})$ might decrease too slow as $\epsilon_{base}$ increases; in other words, $\epsilon_{base}={U^{\mathbf{\rho}}_{Sam}}^{-1}(v)$ and also the price $\pi=\Pi^{\mathbf{\rho}}_{Sam}(v)$ might decrease too quick as $v$ increases, which allows arbitrage behaviors.
\begin{example}[Arbitrage: Sample Mechanism]
\label{arbitrage}
Let $c_i=c$ for all $i$, $A=(1+r)\cdot c$ and $\mathbf{\rho}=[0.5,0.5,...,0.5,1]$. If $[\epsilon_1,...,\epsilon_{49},\epsilon_{50}]=[1,...,1,2]$, $v=U^{\mathbf{\rho}}_{Sam}(2)=49 \frac{e-1}{e^2-1} (1-\frac{e-1}{e^2-1})+2\cdot(\frac{2}{2})^2=11.634$, and $\pi=(1+r)\cdot\sum_i c \cdot \epsilon_i=51A$; if $[\epsilon_1,...,\epsilon_{49},\epsilon_{50}]=[0.41641,...,0.41641,$
$0.83282]$, then $v=U^{\mathbf{\rho}}_{Sam}(0.83282)=23.268$ but $\pi=(1+r)\cdot\sum_i c \cdot \epsilon_i=21.23691A$. Thus, the buyer can buy two independent query answers with $v=23.268$, and then derive a new query answer with $v=\frac{23.268}{2}=11.634$ by combining the former two answers. Because the total price of the two query answers with $v=23.268$ is equal to $42.47382A$ (lower than the price of the query answer with $v=11.634$), arbitrage behaviors are allowed in this case.

\end{example}
\vspace{-4pt}
Unfortunately, since the inverse function ${U^{\mathbf{\rho}}_{Sam}}^{-1}$ is much complicated, it is difficult to prove the arbitrage-freeness of $\Pi^{\mathbf{\rho}}_{Sam}(v)$. Thus, we propose Theorem \ref{Arbitrage-free} to help to prove the arbitrage-freeness, where the first condition makes the pricing function decreasing, the second guarantees a smooth decreasing speed, and the third means a zero price corresponds to infinitely high variance.

\begin{theorem}
\label{Arbitrage-free}
The pricing function $\pi=\Pi^{\mathbf{\rho}}_{\mathcal{M}}(v)$ is arbitrage-free, if the utility function $v=U^{\mathbf{\rho}}_{\mathcal{M}}(\epsilon_{base})$ satisfies the following conditions:
\begin{enumerate}
\item decreasing, which means $\forall \epsilon_{base}>0$, ${U^{\mathbf{\rho}}_{\mathcal{M}}}^{(1)}(\epsilon_{base})< 0$;
\item $\forall \epsilon_{base}>0$, $U^{\mathbf{\rho}}_{\mathcal{M}}(\epsilon_{base})\cdot {U^{\mathbf{\rho}}_{\mathcal{M}}}^{(2)}(\epsilon_{base})\leq 2\cdot [{U^{\mathbf{\rho}}_{\mathcal{M}}}^{(1)}(\epsilon_{base})]^2$;
\item $\lim_{\epsilon_{base} \rightarrow 0^+} U^{\mathbf{\rho}}_{\mathcal{M}}(\epsilon_{base}) = +\infty$.
\end{enumerate}
\end{theorem}

We propose the $N$-Grouping algorithm to allocate privacy budgets such that the pricing function $\Pi^{\mathbf{\rho}}_{Sam}$ is arbitrage-free. Simply speaking, since there might be multiple values of $\mathbf{\rho}$ resulting in arbitrage-freeness, $N$-Grouping finds a pattern $\mathbf{\rho}$ by mathematical optimization where the elements $\rho_i$ can be grouped into $N$ sections by their values, and it allocates privacy budgets in such $\mathbf{\rho}$-pattern. 

Based on Theorem \ref{Arbitrage-free}, $N$-Grouping also converts the task of finding an arbitrage-free pricing function into finding the $\mathbf{\rho}$ such that: $\min\limits_{\mathbf{\rho} \in {R^+}^n}\|\mathbf{\rho}-\mathbf{\rho}^{init}\|_{1}$, s.t. $\forall \epsilon >0$, ${U^{\mathbf{\rho}}_{Sam}}^{(1)}(\epsilon)< 0$ and $U^{\mathbf{\rho}}_{Sam}(\epsilon)\cdot {U^{\mathbf{\rho}}_{Sam}}^{(2)}(\epsilon)\leq 2\cdot [{U^{\mathbf{\rho}}_{Sam}}^{(1)}(\epsilon)]^2$, which is a \textbf{semi-infinite programming} problem. That is, $N$-Grouping finds $\mathbf{\rho}$ by minimizing the Manhattan distance between $\mathbf{\rho}$ and $\mathbf{\rho}^{init}$, subject to the first two sufficient conditions in Theorem \ref{Arbitrage-free}. $\mathbf{\rho}^{init}$ is a initial pattern pre-defined by some function. To utilize more privacy losses, we set $\mathbf{\rho}^{init} = [\hat{\epsilon}_1,...,\hat{\epsilon}_n]/ max_i \hat{\epsilon}_i$ to make $\mathbf{\rho}$ similar to the pattern of data owners' maximum tolerable privacy losses. As depicted in Alg. \ref{Grouping}, $N$-Grouping allocates privacy budgets in the following steps:
\begin{itemize}[leftmargin=*]
	 \item 1-3: Calculate $step$ to divide groups; get the initial pattern $\mathbf{\rho}^{init}$ based on the maximum tolerable privacy losses $[\hat{\epsilon}_1,...,\hat{\epsilon}_n]$; sort $\mathbf{\rho}^{init}$ in ascending order of value temporarily. 
	\item 4-6: Divide the elements of $\mathbf{\rho}^{init}$ into $N$ groups; for each group, make all the values of its members equal to the minimum one. 
	\item 7-9: Recover the order of elements in $\mathbf{\rho}^{init}$ by their corresponding data owners' identities; process semi-infinite non-linear programming; allocate half of the remaining tolerable privacy losses $[\hat{\epsilon}^k_1,...,\hat{\epsilon}^k_n]$ as the initial privacy budgets.
	\item 10-15: Shave the base privacy budget $\epsilon_{base}$. Because there might be cases where some privacy budget $\bar{\epsilon}^k_i$ is lower than $\epsilon_{base} * \rho_i$, we have to lower $\epsilon_{base}$ so that the initial privacy budgets can cover the privacy budgets $\epsilon_{base}*\mathbf{\rho}$.
	\item 16: Output $\epsilon_{base}*\mathbf{\rho}$ as privacy budgets.

\end{itemize}

\begin{algorithm}[h]
\algsetup{linenosize=\small} 
\small
\caption{$N$-Grouping}
\begin{algorithmic}[1]
\REQUIRE $[\hat{\epsilon}^k_1,...,\hat{\epsilon}^k_n]$, $[\hat{\epsilon}_1,...,\hat{\epsilon}_n]$
\ENSURE $[\bar{\epsilon}^k_1,...,\bar{\epsilon}^k_n]$

\STATE $step \gets n / N$; //N should be constant for all queries over $D$.
\STATE $\mathbf{\rho}^{init} \gets [\hat{\epsilon}_1,...,\hat{\epsilon}_n]/ max_i \hat{\epsilon}_i$;
\STATE Sort $\mathbf{\rho}^{init}$ by value;
\FOR {$j=1$ \TO $N$}
	\STATE $\rho^{init}_{(j-1)\cdot step+1},...,\rho^{init}_{j\cdot step} \gets min(\rho^{init}_{(j-1)\cdot step+1},...,\rho^{init}_{j\cdot step})$;
\ENDFOR
\STATE Sort $\mathbf{\rho}^{init}$ by key (data owner's ID);
\STATE Find $\mathbf{\rho}$ by semi-infinite programming;

\STATE $[\bar{\epsilon}^k_1,...,\bar{\epsilon}^k_n] \gets \frac{1}{2}[\hat{\epsilon}^k_1,...,\hat{\epsilon}^k_n]$;
\STATE $\epsilon_{base} \gets \max\limits_i \bar{\epsilon}^k_i$;
\FOR {$i=1$ \TO $n$}
	\IF {$\bar{\epsilon}^k_i < \epsilon_{base} * \rho_i$}
		\STATE $\epsilon_{base} \gets \bar{\epsilon}^k_i/\rho_i$;
	\ENDIF
\ENDFOR
\RETURN $[\bar{\epsilon}^k_1,...,\bar{\epsilon}^k_n] \gets \epsilon_{base}*\mathbf{\rho}$;
\end{algorithmic}
\label{Grouping}
\end{algorithm}

Each of the Laplce-base mechanism and Sample-base mechanism has its own advantages. On the one hand, the Sample-base mechanism is sensitive to the size of database $n$. We note that the output of the utility function $U_{Sam}$ is sensitive to the $n$ while the output of $U_{Lap}$ only depends on $\min_i \epsilon_i$, i.e., the minimum privacy loss among data owners. Thus, even if the privacy budgets are uniform, with the size of database increasing, the performance of the Sample-base mechanism in the aspect of utility is becoming worse while the Laplace-base mechanism remains the same. On the other hand, the Sample-base mechanism can provide lower variance and make more use of the maximum tolerable privacy losses. Under the Sample-base mechanism, as each data owner's privacy loss is able to be personalized, a lower minimum affordable variance $\check{v}$ can be provided, so that the buyer can have more space of choices. Also, since the Laplace-base mechanism computes uniform privacy losses, a portion of each owner's maximum tolerable privacy loss might be waste; meanwhile, under the Sample-base mechanism, such portion of privacy loss can be utilized so that data owners can gain more compensations.

\bibliographystyle{ACM-Reference-Format}
\bibliography{sfdi2019+}

\end{document}